\documentclass[prd,aps,showpacs,preprintnumbers,amssymb]{revtex4}
\usepackage{axodraw}
\usepackage{color}
\usepackage{epsf}

\def\e3p{$\eta \rightarrow 3 \pi$}

\begin{document}
\title{%
\hfill{\normalsize\vbox{%
\hbox{}
 }}\\
{ A semiclassical approach for the Higgs boson }}

\author{Amir H. Fariborz
$^{\it \bf a}$~\footnote[1]{Email:
 fariboa@sunyit.edu}}
\author{Renata Jora
$^{\it \bf b}$~\footnote[2]{Email:
 rjora@theory.nipne.ro}}

\author{Joseph Schechter
 $^{\it \bf c}$~\footnote[3]{Email:
 schechte@phy.syr.edu}}

\affiliation{$^{\bf \it a}$ Department of Mathematics and Physics,  State University of New York Polytechnic Institute, Utica, NY 13502, USA}
\affiliation{$^{\bf \it b}$ National Institute of Physics and Nuclear Engineering PO Box MG-6, Bucharest-Magurele, Romania}

\affiliation{$^ {\bf \it c}$ Department of Physics,
 Syracuse University, Syracuse, NY 13244-1130, USA}

\date{\today}

\begin{abstract}
Starting from the equations of motion of the fields in a theory with spontaneous symmetry breaking and by making some simple assumptions regarding their behavior we derive simple tree level relations between the mass of the Higgs boson in the theory and the masses of the gauge bosons corresponding to the broken generators. We show that these mass relations have a clear meaning if both the scalars and the gauge bosons in the theory are composite states made of two fermions.
\end{abstract}
\pacs{12.60.Cn, 12.60.Fr, 11.15.Ex}
\maketitle

\section{Introduction}

Recent experimental results from the ATLAS \cite{Atlas} and CMS \cite{CMS} experiments at CERN  confirmed not only the existence of a scalar Higgs particle with a mass of $m_H=125.5$ GeV but also that its most important couplings with the fermions are those of the standard model. It is known that the standard model \cite{Glashow}-\cite{Veltman} by itself does not predict the mass of the Higgs boson. However this mass can be obtained if one considers additional assumptions  beside the standard model content \cite{Wetterich}-\cite{Jora1}. Note that the standard model is not only  a gauge theory with spontaneous symmetry breaking but also one of the few theories where the number of scalars coincides with that of the gauge symmetries generators. Of course by spontaneous symmetry breaking the number of massive gauge states should be the same as the number of the Goldstone bosons. In \cite{Jora2} we suggested that the equality of scalar and gauge boson states is not a coincidence and that also a massive scalar field although non-interacting at tree level with a massless gauge boson can be associated  to it through a new effective symmetry of the spontaneously broken Lagrangian. If the scalar and the gauge boson  in a theory are so closely interrelated we expect that somehow also their masses are closely connected.
This is the possibility that we will explore in this short report by considering a semiclassical approach to the Higgs boson of general spontaneously broken gauge theory. Thus we will show by making a few reasonable assumptions that one can obtain simple mass relation between the mass of the Higgs boson and the masses  of the gauge fields corresponding to the
spontaneously broken generators. This approach can be easily applied to more complicated theories as those with  multiple Higgses and with some assumptions about the relevant scales even to supersymmetry. We will only suggest how these results acquire a particular significance in the case of models with strong dynamics and composite particles because  this issue alone deserves a treatment in a separate work.

\section{An abelian Higgs model}
Let us consider the Lagrangian:
\begin{eqnarray}
{\cal L}=-\frac{1}{4}F_{\mu\nu}F^{\mu\nu}+(D^{\mu}\Phi)^{*}(D_{\mu}\Phi)-V(\Phi),
\label{lagr5467}
\end{eqnarray}
where,
\begin{eqnarray}
V(\Phi)=-\mu^2\Phi^*\Phi+\frac{\lambda}{2}(\Phi^*\Phi)^2.
\label{pot7689}
\end{eqnarray}
This potential displays spontaneous symmetry breakdown for a vacuum expectation value of the Higgs boson:
\begin{eqnarray}
\langle \Phi\rangle= \left(\frac{\mu^2}{\lambda}\right)^{\frac{1}{2}}
\label{vev4567}
\end{eqnarray}

The field $\Phi$ is then expanded around the minima as,
\begin{eqnarray}
\Phi=v+(\Phi_1(x)+i\Phi_2(x)),
\label{rez5678}
\end{eqnarray}
where $\Phi_1(x)$ corresponds to the massive scalar boson and $\Phi_2(x)$ to the Goldstone boson eaten by the gauge field $A_{\mu}$ which acquires a mass, $m_A^2=2e^2v^2$. Note that for the sake of simplicity we did not include the usual factor of normalization of $\frac{1}{\sqrt{2}}$ in front of  $\Phi_1$ and $\Phi_2$.

We rewrite the Lagrangian in Eq. (\ref{lagr5467}) in the unitary gauge:
\begin{eqnarray}
{\cal L}=-\frac{1}{4}F^{\mu\nu}F_{\mu\nu}+(\partial_{\mu} \Phi)^2+e^2\Phi^2A_{\mu}A^{\mu}-V(\Phi),
\label{newl789}
\end{eqnarray}
where,
\begin{eqnarray}
V(\Phi)=-\mu^2\Phi^2+\frac{\lambda}{2}\Phi^4,
\label{pot768}
\end{eqnarray}
and the new $\Phi$ is the real part of the initial field that develops a vacuum expectation value and that is identified with the Higgs boson.

We next determine the equations of motion for the fields  $\Phi$ and $A_{\mu}$ from the Lagrangian in Eq. (\ref{newl789}):
\begin{eqnarray}
&&\partial_{\mu}\partial^{\mu}\Phi-\mu^2\Phi+\lambda\Phi^3-2e^2\Phi A^{\mu}A_{\mu}=0
\nonumber\\
&&(-\partial^2 g^{\mu\nu}+\partial^{\mu}\partial^{\nu})A_{\mu}-2e^2\Phi^2A^{\nu}=0.
\label{mot12879}
\end{eqnarray}
The standard procedure is to extract from the system in Eq. (\ref{mot12879}) the free field propagators and add the rest of the terms in the functional approach as interaction terms.
We shall use a different procedure here namely we solve for $\Phi$ from the equation of motion for $A_{\nu}$  to obtain:
\begin{eqnarray}
\Phi=\pm\sqrt{\frac{(-\partial^2 g_{\mu\nu}+\partial_{\mu}\partial_{\nu})A^{\mu}}{2e^2A_{\nu}}}
\label{field6578}
\end{eqnarray}
We next separate the gauge field in a slowly varying background gauge field which we force to satisfy the on-shell equation of motion (Note that we can pick any gauge condition for the background gauge field),
\begin{eqnarray}
-\partial^2B_{\nu}-m_A^2B_{\nu}=0
\label{motion6789}
\end{eqnarray}
and a fluctuation $\tilde{A}_{\mu}$ such that $A_{\mu}=B_{\mu}+\tilde{A}_{\mu}$.  Then Eq. (\ref{field6578}) can be simplified as :
\begin{eqnarray}
\Phi=\pm\left(\frac{B_{\nu}+[\frac{1}{m_A^2}(-\partial^2g_{\mu\nu}+\partial_{\mu}\partial_{\nu})\tilde{A}^{\mu}]}{A_{\nu}}\right)^{1/2}.
\label{rez67589}
\end{eqnarray}
This result can be regarded from several points of view. We are interested in a tree level relation satisfied  by the Higgs boson.  If we consider the field $\Phi$ as a small expansion around the vev (we set overall $v=1$ for the convenience of the calculations) which is equivalent to the expansion  in the small parameter $\frac{\tilde{A_{\nu}}}{B_{\mu}}$ we can write in first order:
\begin{eqnarray}
1+h=1-\frac{1}{2}\frac{1}{B_{\nu}}\left[\frac{1}{m_A^2}(-\partial^2g_{\mu\nu}+\partial_{\mu}\partial_{\nu})\tilde{A}^{\mu}-\tilde{A}_{\nu}\right]=1-\frac{1}{2}\frac{D_{\mu\nu}\tilde{A}^{\mu}}{B_{\nu}},
\label{rez4356}
\end{eqnarray}
where we denote,
\begin{eqnarray}
\frac{1}{m_A^2}(-\partial^2g_{\mu\nu}+\partial_{\mu}\partial_{\nu})\tilde{A}^{\mu}-\tilde{A}_{\nu}=D_{\mu\nu}\tilde{A}^{\mu}.
\label{not5678}
\end{eqnarray}

We then apply the operator $\partial^2$ to both sides of the Eq. (\ref{rez4356}) to obtain:
\begin{eqnarray} -4\mu^2h=-3m_A^2h+{\rm nonlinear\,terms},
\label{firstrez5467}
\end{eqnarray}
where on the left hand side we applied the equation of motion for the Higgs field and on the right hand side  we used:
\begin{eqnarray}
&&\partial^{\rho}\partial_{\rho}\frac{D_{\mu\nu}\tilde{A}^{\mu}}{B_{\nu}}=
3\frac{\partial^{\rho}\partial_{\rho}D_{\mu\nu}\tilde{A}^{\mu}}{B_{\nu}}-
3\frac{\partial^{\rho}\partial_{\rho}B_{\nu}D_{\mu\nu}\tilde{A}^{\mu}}{B_{\nu}^2}
\nonumber\\
&&-2\frac{1}{B_{\nu}^2}\partial^{\rho}[B_{\nu}\partial_{\rho}D_{\mu\nu}\tilde{A}^{\mu}]+
2\frac{1}{B_{\nu}^3}D_{\mu\nu}\tilde{A}^{\mu}\partial^{\rho}[B_{\nu}\partial_{\rho}B_{\nu}].
\label{rez435627}
\end{eqnarray}

Since we are interested only in linear terms solution to an eigenvalue problem the last two terms in the second  line of the Eq. (\ref{rez435627}) can be neglected. The crucial point is that we can pick the background gauge field as we want and the result should be independent of our choice. If we tune the background gauge field instead to be a large but slowly varying field which does not necessarily satisfy the equation of motion of a massive particle we would obtain that  the mass of the fluctuation $\tilde{A}_{\nu}$ is just the mass of the Higgs.  This is natural also by a simple inspection of the equation that defines the fluctuation of the scalar with respect to the vacuum. Then  the transition from Eq. (\ref{rez435627}) to Eq. (\ref{firstrez5467}) is straightforward.

There is an alternative approach to  the above reasoning that applies to the big deviation of the gauge field from the background gauge field. In this case the field $\tilde{A}_{\mu}$ can be considered large and almost constant whereas the background gauge field satisfies the equation of motion of a massive gauge boson in the Feynman gauge with a mass $m_A^2$. We set to zero the fluctuations of $\tilde{A}_{\mu}$ as corresponding only to loop corrections. This yields,
\begin{eqnarray}
\Phi^2=\frac{B_{\mu}}{A_{\mu}}
\label{new54678}
\end{eqnarray}

We further apply the operator $\partial^2$ to both sides of the Eq. (\ref{new54678}) and look for solutions to the eigenvalue equation:
\begin{eqnarray}
&&2\Phi\partial^{\rho}\partial_{\rho}\Phi+2\partial^{\rho}\Phi\partial_{\rho}\Phi=\frac{3\partial^{\rho}\partial_{\rho}B_{\mu}}{A_{\mu}}-
\nonumber\\
&&2\frac{\partial^{\rho}(A_{\mu}\partial_{\rho}B_{\mu})}{A_{\mu}^2}-
\frac{3B_{\mu}\partial^{\rho}\partial_{\rho}A_{\mu}}{A_{\mu}^2}+\frac{2B_{\mu}\partial^{\rho}A_{\mu}\partial_{\rho}A_{\mu}}{A_{\mu}^2}
\label{rez54678}
\end{eqnarray}

Then one obtains by considering the terms which are only quadratic in field (and regard the rest as interactions):
\begin{eqnarray}
-2(2\mu^2)\Phi^2+{\rm nonlinear\,terms}=-3m_A^2\frac{B_{\mu}}{A_{\mu}}+{\rm nonlinear\,terms}=-3m_A^2\Phi^2+{\rm nonlinear\,terms}
\label{tr6578}
\end{eqnarray}

Finally form two point of view using reasonable assumptions we obtain the following mass relation in a spontaneously broken abelian gauge theory:
\begin{eqnarray}
2m_h^2=3m_A^2,
\label{fin678}
\end{eqnarray}
where $m_A$ is the mass of the vector boson and $m_h$ is the mass of the corresponding Higgs boson.
\section{Tree level relations in the standard model}

We shall apply the procedure in the previous  section to the standard model. Let us consider the Higgs gauge Lagrangian of the standard model after spontaneous symmetry breaking in the unitary gauge (as taken from \cite{Mark}):
\begin{eqnarray}
&&{\cal L}=-\frac{1}{4}F^{\mu\nu}F_{\mu\nu}-\frac{1}{4}Z^{\mu\nu}Z_{\mu\nu}-D^{\dagger\mu}W^{-\mu}D_{\mu}W_{\nu}+D^{\dagger\mu}W^{-\nu}D_{\nu}W_{\mu}^{+}+
\nonumber\\
&&+ie(F^{\mu\nu}+\cot\theta_WZ^{\mu\nu})W_{\mu}^{+}W_{\nu}^{-}
\nonumber\\
&&-\frac{1}{2}\frac{e^2}{\sin^2\theta_W}(W^{+\mu}W_{\mu}^{-}W^{+\nu}W_{\nu}^{-}-W^{+\mu}W_{\nu}^{+}W^{-\nu}W_{\nu}^{-})
\nonumber\\
&&(-m_W^2W^{+\mu}W_{\mu}^{-}+\frac{1}{2}M_Z^2Z^{\mu}Z_{\nu})(1+v^{-1}h)^2-\frac{1}{2}\partial^{\mu}h\partial_{\mu}h-V(h).
\label{rezt546789}
\end{eqnarray}

We will denote $\Phi=1+v^{-1}h$ for the rest of the calculations and also set $v=1$. Note that the above Lagrangian does not contain fermion fields. One could consider also the fermion fields or even only the fermion fields case in which different mass relations could be obtained. However we will try to follow the same line of reasoning as for the abelian gauge field and consider the fermion fields only as interaction terms and ignore them for the rest of the calculations.

The next step is to determine the equations of motion for the Z:
\begin{eqnarray}
&&(-\partial^2g_{\mu\nu}+\partial_{\mu}\partial_{\nu})Z^{\nu}+
\nonumber\\
&&ie\cot\theta_W\left[\partial_{\mu}W_{\nu}^{-}W^{+\nu}
-\partial_{\mu}W_{\nu}^{+}W^{-\nu}-\partial_{\nu}W_{\mu}^{-}W^{+\nu}+\partial_{\nu}W_{\mu}{+}W^{-\nu}\right]
\nonumber\\
&&+ie\cot\theta_W \partial_{\nu}(W_{\mu}^{+}W^{-\nu})-m_Z^2\Phi^2Z_{\mu}=0,
\label{zer789}
\end{eqnarray}
and W bosons:
\begin{eqnarray}
&&(\partial^2g_{\mu\nu}-\partial_{\mu}\partial_{\nu})W^{+\nu}+ie\partial_{\nu}[(A^{\nu}+\cot\theta_WZ^{\nu})W^{-\mu}-(A_{\mu}+\cot\theta_W Z_{\mu})W^{+\nu}]
\nonumber\\
&&+ie F_{\mu\nu}W^{+\nu}-ie\cot\theta_W Z_{\mu\nu}W^{+\nu}+
\frac{e^2}{\sin^2\theta_W}[W^{+\nu}W^{-}_{\nu}W^{+}_{\mu}-W^{+\nu}W^{+}_{\nu}W^{-}_{\mu}]-m_W^2\Phi^2W^{+}_{\mu}]=0.
\label{wb8908}
\end{eqnarray}

From Eqs. (\ref{zer789}), (\ref{wb8908}) one can extract two solutions for the field $\Phi^2$:

\begin{eqnarray}
&&\Phi^2=[(-\partial^2g_{\mu}{\nu}+\partial_{\mu}\partial_{\nu})Z^{\nu}+
\nonumber\\
&&ie\cot\theta_W\left[\partial_{\mu}W_{\nu}^{-}W^{+\nu}
-\partial_{\mu}W_{\nu}^{+}W^{-\nu}-\partial_{\nu}W_{\mu}^{-}W^{+\nu}+\partial_{\nu}W_{\mu}{+}W^{-\nu}\right]
\nonumber\\
&&+ie\cot\theta_W \partial_{\nu}(W_{\mu}^{+}W^{-\nu}]/[m_Z^2Z_{\mu}]
\nonumber\\
&&\Phi^2=[W^{-\mu}[(\partial^2g_{\mu\nu}-\partial_{\mu}\partial_{\nu})W^{+\nu}+ie\partial_{\nu}[(A^{\nu}+\cot\theta_WZ^{\nu})W^{-\mu}-(A_{\mu}+\cot\theta_W Z_{\mu})W^{+\nu}]
\nonumber\\
&&-ie F_{\nu\mu}W^{+\nu}-ie\cot\theta_W Z_{\mu\nu}W^{+\nu}+
\frac{e^2}{\sin^2\theta_W}[W^{+\nu}W^{-}_{\nu}W^{+}_{\mu}-W^{+\nu}W^{+}_{\nu}W^{-}_{\mu}]]/[m_W^2W^{+}_{\mu}W^{-\mu}.
\label{rez6578490}
\end{eqnarray}
In the last expression for $\Phi^2$ we multiplied the numerator and the denominator by $W^{-\mu}$(Note that there is no summation over the index $\mu$).
Eq. (\ref{rez6578490}) seems very complicated to deal with.  In order to simplify it we first decompose the vector boson fields in a background gauge field and a fluctuation:
\begin{eqnarray}
&&Z_{\mu}=Z_{\mu}'+\tilde{Z}_{\mu}
\nonumber\\
&&W^{+}_{\mu}=W^{+\prime}_{\mu}+\tilde{W}^{+}_{\mu}.
\label{rez56478}
\end{eqnarray}
such that the fields $Z_{\mu}'$ and $W^{+\prime}_{\mu}$ satisfy the equations:
\begin{eqnarray}
&&-\partial^2Z_{\mu}'-m_Z^2Z_{\mu}'=0
\nonumber\\
&&-\partial^2W^{+\prime}_{\mu}-m_W^2W^{+\prime}_{\mu}=0.
\label{mot879}
\end{eqnarray}

Even in this case the  solutions  for $\Phi^2$ are too complicated and one needs to eliminate all the interaction terms in order to get an eigenvalue equation.  This means that the result we obtain can be regarded only as a first order approximation. One then can apply the procedure of section II to either of the solutions in the Eq. (\ref{rez6578490}) to obtain a mass relation which is only a rough approximation:
\begin{eqnarray}
&&3m_Z^2\approx 2m_h^2
\nonumber\\
&&3m_W^2\approx 2m_h^2.
\label{mass54678}
\end{eqnarray}
Depending on which solutions one uses operators suppressed by $2m_W^2$ or $m_Z^2$ are neglected. For example a typical interaction term which contains two W' in the numerator in the equation for Z will bring upon applying the equation of motion a suppression of $2m_W^2/m_Z^2$ which is larger than one (this besides the extra small factors that are in front which together still justify the neglect of this terms). However since we are working with parameters at the electroweak scale such suppression can lead only to rough estimates. One particular choice for the solution for $\Phi^2$ minimizes the contribution of most interaction terms and also takes into account all the gauge bosons. This is:
\begin{eqnarray}
\Phi=\left[\frac{Z_{\mu}^{\prime}}{Z_{\mu}}\frac{W^{+\nu\prime}W^{-\prime}_{\nu}}{W^{+}_{\nu}W^{-\nu}}\right]^{1/4}+{\rm suppressed\,terms\,that\,are\,neglected}
\label{finalexpr56789}
\end{eqnarray}
Then one applies the second procedure in section II (which is more suitable) where terms with two partial derivatives are neglected if they cannot lead to an operator $\partial^2$ applied to the same operator by integration by parts. The calculations lead to the following mass relation:
\begin{eqnarray}
3(m_Z^2+2m_W^2)\approx 4m_h^2,
\label{rez54678}
\end{eqnarray}
which predicts a mass of the standard model Higgs boson of $m_h=126.21$ GeV. Note that it is a valid prediction for the assumptions  sketched in section II.

\section{Conclusions}

In this work we considered the equation of motions of the Higgs boson and those of the gauge fields in a  theory with spontaneous symmetry breakdown and  considered the behavior of these solutions for two cases, when the fluctuation of the gauge field is much smaller that the background gauge field or much bigger to predict a mass relation which connects the masses of the Higgs bosons with those of the gauge bosons. Note that for different Anstaze different relations can be obtained which  are better or worse approximations depending on the choice of the solutions we use. The method discussed is only approximate and consists in the systematic elimination of some possible terms based on the initial hypothesis regarding the behavior of the fields involved.  As the complexity of the theory increases this procedure requires the introduction of more fields  related to the Higgs boson  in order to make a sufficiently reliable approximation. It turns out that for one particular choice given in Eq. (\ref{finalexpr56789}) one can predict a mass of the Higgs boson of $m_h=126.21$ GeV which is in excellent agreement with the experimental value.

However there is a situation where  relations as those in Eqs. ({\ref{fin678}) and (\ref{rez54678}) can be considered exact.  We shall illustrate this for the simple case of an abelian gauge theory with the Lagrangian given in Eq. (\ref{lagr5467}).
The situation of interest is when both the scalar and the gauge boson in the model are composites of the same particle according to:
\begin{eqnarray}
&&\Phi=\bar{\Psi}\Psi+\bar{\Psi}\gamma^{5}\Psi
\nonumber\\
&&A_{\mu}=\bar{\Psi}\gamma_{\mu}\Psi.
\label{composi678}
\end{eqnarray}
Consider the abelian gauge theory in the unitary gauge after spontaneous symmetry breakdown. The mass term for the Higgs boson in terms of the composite states is given by:
\begin{eqnarray}
{\cal L}_m=-\frac{m_h^2}{M^2}\bar{\Psi}\Psi\bar{\Psi}\Psi,
\label{rez43567}
\end{eqnarray}
where $M$ is the composite scale.  We apply  the following Fierz transformation to the right hand side of Eq. (\ref{rez43567}):
\begin{eqnarray}
&&\bar{\Psi}\Psi\bar{\Psi}\Psi=\frac{1}{4}\bar{\Psi}\Psi\bar{\Psi}\Psi+\frac{1}{4}\bar{\Psi}\gamma^{\mu}\Psi\bar{\Psi}\gamma_{\mu}\Psi+\frac{1}{8}\bar{\Psi}\sigma^{\rho\sigma}\Psi\bar{\Psi}\sigma_{\rho\sigma}\Psi-
\nonumber\\
&&-\frac{1}{4}\bar{\Psi}\gamma^{\nu}\gamma^5\Psi\bar{\Psi}\gamma_{\nu}\gamma^5\Psi+\frac{1}{4}\bar{\Psi}\gamma^5\Psi\bar{\Psi}\gamma^5\Psi,
\label{fin657892}
\end{eqnarray}
and then successively to the scalar part in Eq. (\ref{fin657892}). We also neglect the other quark bilinears to get (using $\sum_{n=1}\frac{1}{4^n}=\frac{1}{3}$):
\begin{eqnarray}
&&\bar{\Psi}\Psi\bar{\Psi}\Psi=\frac{1}{3}\bar{\Psi}\gamma^{\mu}\Psi\bar{\Psi}\gamma_{\mu}\Psi+....
\nonumber\\
&&\frac{m_h^2}{M^2}\bar{\Psi}\Psi\bar{\Psi}\Psi=\frac{m_h^2}{M^2}\frac{1}{3}\bar{\Psi}\gamma^{\mu}\Psi\bar{\Psi}\gamma_{\mu}\Psi+...=
\frac{m_A^2}{2}\frac{1}{M^2}\bar{\Psi}\gamma^{\mu}\Psi\bar{\Psi}\gamma_{\mu}\Psi+...,
\label{rez4567}
\end{eqnarray}
which yields $3m_A^2=2m_h^2$. (Note that one needs to work in a metric in which both mass terms for the Higgs and gauge boson have the same sign). We thus claim that the same term in the composite Lagrangian is responsible for both masses.

There is a more transparent way to see this.
Assume we have a composite Lagrangian of the Nambu-Jona-Lasinio type with a four fermion interaction which leads to condensation  and to spontaneous symmetry breakdown. The relevant four fermion interaction is given by:
\begin{eqnarray}
{\cal L}=\frac{m_h^2}{M^2}(\bar\Psi\Psi-\bar\Psi\gamma^{5}\Psi)(\bar\Psi\Psi+\bar\Psi\gamma^{5}\Psi)=
\frac{m_h^2}{M^2}\bar\Psi\Psi\bar\Psi\Psi-\frac{m_h^2}{M^2}\bar\Psi\gamma^{5}\Psi\bar\Psi\gamma^{5}\Psi.
\label{rer657}
\end{eqnarray}
There is no need of quartic scalar interaction in this Lagrangian. The first term gives the mass to the Higgs boson whereas the second converts into the mass of the gauge boson in the following way.
 We apply the following Fierz transformation to the pseudoscalar term:
\begin{eqnarray}
&&\bar{\Psi}\gamma^5\Psi\bar{\Psi}\gamma^5\Psi=\frac{1}{4}\bar{\Psi}\Psi-\frac{1}{4}\bar{\Psi}\gamma^{\mu}\Psi\bar{\Psi}\gamma_{\mu}\Psi+
\frac{1}{8}\bar{\Psi}\sigma^{\rho\sigma}\Psi\bar{\Psi}\sigma_{\rho\sigma}\Psi-
\nonumber\\
&&+\frac{1}{4}\bar{\Psi}\gamma^{\nu}\gamma^5\Psi\bar{\Psi}\gamma_{\nu}\gamma^5\Psi+\frac{1}{4}\bar{\Psi}\gamma^5\Psi\bar{\Psi}\gamma^5\Psi,
\label{fin65789}
\end{eqnarray}
and again an infinite number of times to the pseudoscalar term in the above relation to get:
\begin{eqnarray}
\bar{\Psi}\gamma^5\Psi\bar{\Psi}\gamma^5\Psi=-\frac{1}{3}\bar{\Psi}\gamma^{\mu}\Psi\bar{\Psi}\gamma_{\mu}\Psi+{\rm \,other\,terms}.
\label{res546777}
\end{eqnarray}
Then a mass term of $m_h^2$ on both sides of this equation leads to $3m_A^2=2m_h^2$ where the mass term for the pseudoscalar has been converted into a mass term for the gauge boson.
This argument is quite general and can apply as well to composite vector, scalar and pseudoscalar mesons in QCD and to the standard model itself. However the possibility that not only the Higgs doublet but also all the gauge bosons in the standard model are composite  deserves to be treated in detail in a separate work.

\section*{Acknowledgments} \vskip -.5cm
The work of R. J. was supported by a grant of the Ministry of National Education, CNCS-UEFISCDI, project number PN-II-ID-PCE-2012-4-0078.

\end{document}